\documentclass[12pt]{article}        
\begin{document}
\def\p {{\partial}}
\def\n {{\nu}}
\def\m {{\mu}}
\def\a {{\alpha}}
\def\bt {{\beta}}
\def\f {{\phi}}
\def\th {{\theta}}
\def\g {{\gamma}}
\def\eps {{\epsilon}}

\def\e {{\psi}}
\def\la {{\lambda}}
\def\na {{\nabla}}
\def\k {\chi}
\def\p {{\partial}}
\def\n {{\nu}}
\def\m {{\mu}}
\def\a {{\alpha}}
\def\bt {{\beta}}
\def\f {{\phi}}
\def\th {{\theta}}
\def\g {{\gamma}}
\def\eps {{\epsilon}}

\def\e {{\psi}}
\def\la {{\lambda}}
\def\na {{\nabla}}
\def\k {\chi}

\begin{center}
Fractional Hamilton formalism within Caputo's derivative
\end{center}

\begin{center}
{ Dumitru Baleanu\footnote[1]{On leave of absence from Institute
of Space Sciences, P.O.BOX, MG-23, R 76900, Magurele-Bucharest,
Romania,E-mails: dumitru@cankaya.edu.tr, baleanu@venus.nipne.ro}\\
Department of Mathematics and Computer Sciences, Faculty of Arts
and Sciences, \c{C}ankaya University- 06530, Ankara, Turkey\\
Om. P. Agrawal\\
 Southern Illinois University Carbondale, Illinois
- 62901, USA}
\end{center}
\begin{abstract}
In this paper we develop a fractional Hamiltonian formulation for
dynamic systems defined in terms of fractional Caputo derivatives.
Expressions for fractional canonical momenta and fractional
canonical Hamiltonian are given, and a set of fractional
Hamiltonian equations are obtained.  Using an example, it is shown
that the canonical fractional Hamiltonian and the fractional
Euler-Lagrange formulations lead to the same set of equations.

\end{abstract}

\section{Introduction}

Fractional calculus is one of the generalizations of the classical
calculus and it has been used successfully in various fields of
science and engineering [1-10].
 A huge amount of mathematical knowledge on fractional integrals and
derivatives has been constructed [1-8]. The fractional derivatives
represent the infinitesimal generators of a class of translation
invariant convolution semigroups which appear universally as
attractors.

 During the last decade several
papers, which deal with fractional variational calculus and its
applications, have been published [11-30]. These applications
include classical and quantum mechanics, field theory, and optimal
control. Due the properties of the fractional derivatives the
corresponding theories become non-local and non-commutative. In
the papers cited above, the problems have been formulated mostly
in terms of two types of fractional derivatives, namely
Riemann-Liouville (RL) and Caputo. Among mathematicians, RL
fractional derivatives have been popular largely because they are
amenable to many mathematical manipulations. However, the RL
derivative of a constant is not zero, and in many applications it
requires fractional initial conditions which are generally not
specified. Many believe that fractional initial conditions are not
physical. In contrast, Caputo derivative of a constant is zero,
and a fractional differential equation defined in terms of Caputo
derivatives require standard boundary conditions.  For these
reasons, Caputo fractional derivatives have been popular among
engineers and scientists.

Recently, Agrawal \cite{Agrawalcankaya1} demonstrated that
fractional terminal conditions may be necessary even when a
problem is formulated in terms of Caputo derivatives.  In
\cite{Agrawalcankaya1} it also argued that fractional initial
conditions may have physical meaning. For example, in a one
dimensional heat diffusion process 1/2 order time derivative of
temperature may represent heat flux across a surface.  This blurs
the distinctions between advantages and disadvantages of RL and
Caputo derivatives.  Thus, which derivative would be more suitable
for formulating engineering and scientific problems remain an open
issue.

Recently, the fractional Hamiltonian formulations were presented
for fractional discrete and continuous systems whose dynamics were
defined in terms of RL derivatives [20-27]. In this paper we
construct a fractional Hamiltonian formulation for discrete
systems whose dynamics have been described using Caputo
derivatives.  It is demonstrated in \cite{Agrawalcankaya1} that
the Euler-Lagrange equation of a fractional variational problem
defined in terms of Caputo derivatives only includes both the RL
and the Caputo derivatives. The same is true of the fractional
Hamiltonian formulation discussed here. An example is solved in
detail to demonstrate an application of the Formulation.

\section{Mathematical Tools}
In this section we briefly present some fundamental definitions
used in the previous section.
 The left and the right Riemann-Liouville
and Caputo fractional derivatives are defined as follows:\\

\noindent {\em The Left Riemann-Liouville Fractional Derivative}
\begin{equation}
_aD_x^\alpha f(x) = \frac{1}{\Gamma (n-\alpha)} \left(
\frac{d}{dx} \right)^n \int_a^x (x-\tau)^{n-\alpha-1} f(\tau)
d\tau,
\end{equation} 

\noindent
{\em The Right Riemann-Liouville Fractional Derivative}
\begin{equation}
_xD_b^\alpha f(x) = \frac{1}{\Gamma (n-\alpha)} \left(-\frac{d}{dx}
\right)^n \int_x^b (\tau-x)^{n-\alpha-1} f(\tau) d\tau,
\end{equation} 

The corresponding Caputo's fractional derivatives are defined as
follows\\

 \noindent {\em The Left Caputo Fractional Derivative}
\begin{equation}
_a^CD_x^\alpha f(x)
= \frac{1}{\Gamma (n-\alpha)} \int_a^x (x-\tau)^{n-\alpha-1}
\left( \frac{d}{d\tau} \right)^n f(\tau) d\tau ,
\end{equation} 

\noindent and\\

{\em The Right Caputo Fractional Derivative}
\begin{equation}
_x^CD_b^\alpha f(x) = \frac{1}{\Gamma (n-\alpha)} \int_x^b
(\tau-x)^{n-\alpha-1} \left(-\frac{d}{d\tau} \right)^n f(\tau)
d\tau ,\end{equation} 

\noindent
where $\alpha$ is the order of the derivative such that
$n-1 < \alpha < n$.  These derivatives will be denoted as the
LRLFD, the RRLFD, the LCFD, and the RCFD, respectively.

Our fractional Hamiltonian formulation presented here is based on
the following theorem \cite{Agrawalcankaya1}.  This
theorem is stated here (without proof) for completeness. \\

\noindent
{\bf Theorem } {\em Let $J[q]$ be a functional of the form
\begin{equation}
J[q] = \int_a^b L(t, q, \,_a^CD_t^\alpha q, \,_t^CD_b^\beta q) dt
\end{equation}
\noindent
{\em  where $0 < \alpha, \beta < 1$} and defined on the
set of functions $y(x)$ which have continuous LCFD of order
$\alpha$ and RCFD of order $\beta$ in $[a, b]$. Then a necessary
condition for $J[q]$ to have an extremum for a given function
$q(t)$ is that $q(t)$ satisfy the generalized Euler-Lagrange
equation given by }
\begin{equation}\label{eleq}
\frac{\partial L}{\partial q} + \,_tD_b^\alpha
\frac{\partial L }{\partial \,_a^CD_t^\alpha q } + \,_aD_t^\beta
\frac{\partial L }{\partial \,_t^CD_b^\beta q} = 0, \hspace{0.2in}
t \in [a, \,\, b]
\end{equation} 
\noindent
{\em and the transversality conditions given by}
\begin{equation}
\left[ _tD_b^{\alpha-1} ( \frac{\partial L
}{\partial \,_a^CD_t^\alpha q } ) - _aD_t^{\beta-1} (
\frac{\partial L }{\partial \,_t^CD_b^\beta q } ) \right] \eta(t)
|_a^b = 0.
\end{equation}
\noindent
The proof of the theorem can be found in \cite{Agrawalcankaya1}.

\section{Fractional Hamilton formulation}
In this section we construct the Hamiltonian formulation within
Caputo's fractional derivatives.

 Let us consider the fractional
Lagrangian as given below
\begin{equation}\label{unu}
L(q,_a^CD_t^\alpha q,_t^C D_b^\beta q,t ), \hspace{0.2in}
0<\alpha,\beta< 1.
\end{equation}
By using (\ref{unu}) we define the canonical momenta $p_\alpha$
and $p_\beta$ as follows
\begin{equation}\label{doi}
p_\alpha=\frac{\partial L}{\partial _a^CD_t^\alpha q },
\hspace{0.2in} p_\beta=\frac{\partial L}{\partial _t^CD_b^\beta q
}.
\end{equation}
Making use of (\ref{unu}) and (\ref{doi}) we define the fractional
canonical Hamiltonian as
\begin{equation}\label{trei}
H=p_\alpha \, {_a^CD_t^\alpha q}+{p_{\beta}} \, _t^CD_b^\beta q-
L.
\end{equation}
Taking total differential of (\ref{trei}) and by using
(\ref{doi}), we obtain
\begin{equation}
dH={dp_{\alpha}} \, _a^CD_t^\alpha q + {dp_{\beta}} \,
_t^CD_b^\beta q-\frac{\partial L}{\partial q}dq - \frac{\partial
L}{\partial t}dt.
\end{equation}
Taking into account the fractional Euler-Lagrange equations
(\ref{eleq}) we obtain
\begin{equation}
dH={dp_{\alpha}} \, _a^CD_t^\alpha q + {dp_{\beta}} \, _t^CD_b^\beta q
+(\,_tD_b^\alpha p_{\alpha} + \,_aD_t^\beta p_{\beta})dq -
\frac{\partial L}{\partial t}dt.
\end{equation}
Finally, after some simple manipulations the fractional Hamilton
equations are obtained as follows

\begin{eqnarray}\label{ul}
\frac{\partial H}{\partial t}=-\frac{\partial L}{\partial t},
\hspace{0.1in} \frac{\partial H}{\partial p_{\alpha}}= \,
_a^CD_t^\alpha q, \hspace{0.1in} \frac{\partial H}{\partial
p_{\beta}}= \, _t^CD_b^\beta q, \hspace{0.1in} \frac{\partial
H}{\partial q}= \, _tD_b^\alpha p_{\alpha} + \,_aD_t^\beta
p_{\beta}.
\end{eqnarray}

In the following, an example is considered to demonstrate an
application these equations.

 \subsection{Example}
Consider the following problem \cite{Agrawalcankaya1}: Minimize
\begin{equation}\label{patru}
J[q]= \frac{1}{2} \int_0^1 ( \,_0^CD_t^\alpha q -
\frac{\Gamma(1+\beta)}{\Gamma(1+ \beta-\alpha)}
t^{(\beta-\alpha)})^2 dt, \hspace{0.5in} 0 < \alpha < 1,
\end{equation}
\noindent such that $ q(0) = 0, \mbox{and}~  q(1) = 1$.  For this
example, the Lagrangian is given by
\begin{equation}
L=( \,_0^CD_t^\alpha q - \frac{\Gamma(1+\beta)}{\Gamma(1+
\beta-\alpha)} t^{(\beta-\alpha)})^2.
\end{equation}
The corresponding Euler-Lagrange equation has the form
\begin{equation}\label{ccinci}
\,_tD_1^\alpha ( \,_0^CD_t^\alpha q –
\frac{\Gamma(1+\beta)}{\Gamma(1+ \beta-\alpha)}
t^{(\beta-\alpha)} ) = 0,
\end{equation}
and admits the solution as
\begin{equation}
q(t) = t^\beta.
\end{equation}
The fractional canonical momenta is given by
\begin{equation}
p_{\alpha}=\,_0^CD_t^\alpha q - \frac{\Gamma(1+\beta)}{\Gamma(1+
\beta-\alpha)} t^{(\beta-\alpha)}
\end{equation}
and the fractional canonical Hamiltonian is given by
\begin{equation}
H=p_{\alpha} {_0^CD_t^\alpha} q-L
\end{equation}
or
\begin{equation}
H=\frac{p_{\alpha}^2}{2}+
p_{\alpha}\frac{\Gamma(1+\beta)}{\Gamma(1+ \beta-\alpha)}
t^{(\beta-\alpha)}.
\end{equation}
The canonical equations of motion are given by
\begin{equation}
\frac{\partial H}{\partial p_{\alpha}}= _a^CD_t^\alpha q,
\hspace{0.2in}
\frac{\partial H}{\partial q}= _tD_b^\alpha p_{\alpha}
\end{equation}
or
\begin{equation}\label{cinci}
p_{\alpha}+\frac{\Gamma(1+\beta)}{\Gamma(1+ \beta-\alpha)}
t^{(\beta-\alpha)}=_0^CD_t^\alpha q,_tD_1^\alpha p_{\alpha}=0.
\end{equation}
By inspection we observed that (\ref{cinci}) is equivalent with
(\ref{ccinci}).

\section{Conclusion}

Fractional calculus was intensively applied, especially during the
last decade, for describing the dynamics of the complex systems.
The canonical fractional formulation is still an open problem of
this new and emerging field.

 The characteristic of the constructed fractional Hamilton equations
within Caputo's  derivatives is that both RL and Caputo's
fractional derivatives are involved. Generally, to find a solution
for the system
 (\ref{ul}) we have to replace RL fractional derivative
in terms of Caputo's fractional derivative and then to solve the
corresponding system.

For a given mechanical example we observed that both fractional
Euler-Lagrange equations and fractional Hamiltonian equations give
the same result. The classical results are obtained as a
particular case of the fractional formulation.

The fractional Lagrangians and their corresponding fractional
Hamiltonians are non-local. This property comes from the
definitions of the fractional derivatives, therefore finding the physical interpretation of these derivative is an open issue.\\

{\small The first author would like to thank the organizers of
this colloquium for giving him the opportunity to attend this
meeting. The research reported in this paper was partially
supported by the Scientific and Technical Research Council of
Turkey.}



\begin{thebibliography}{0}            
\bibitem{oldham} K. B. Oldham and J. Spanier: \emph{The Fractional
Calculus}, Academic Press, New-York, (1974).

\bibitem{miller} K. S. Miller and  B. Ross: \emph{An Introduction to the
Fractional Integrals and Derivatives-Theory and Applications},
John Wiley and Sons Inc., New York, (1993).

\bibitem{samko} S. G. Samko, A. A.  Kilbas and O. I.  Marichev:
 \emph{Fractional Integrals and Derivatives - Theory and Applications},
Gordon and Breach, Linghorne, P.A., (1993).

\bibitem{hilfer} R. Hilfer:  \emph{Applications of Fractional Calculus in
Physics},  World Scientific Publishing Company, Singapore, (2000).

\bibitem{podlubny}
I. Podlubny: \emph{Fractional Differential Equations}, Academic
Press, San Diego CA, (1999).
\bibitem{zaslavsky} G. M. Zaslavsky: \emph{Hamiltonian Chaos and Fractional Dynamics},
Oxford University Press, Oxford, (2005).
\bibitem{trujillo}
A. A. Kilbas, H. M. Srivastava and J. J. Trujillo: \emph{Theory
and Applications of Fractional Differential Equations}, Elsevier,
(2006).
\bibitem{richard}
R. L. Magin: \emph{Fractional Calculus in Bioengineering}, Begell
House Publisher, Inc. Connecticut, (2006).
\bibitem{mainardi1}
F. Mainardi, Yu. Luchko and G. Pagnini: \emph{Frac. Calc. Appl.
Analys.} \textbf{4(2)} (2001) 153.
\bibitem{naber}
  M. Naber: J. Math. Phys. \textbf{45} (2004) 3339.

\bibitem{riewe1}
F. Riewe:  Phys. Rev. E \textbf{53} (1996) 1890.
\bibitem{riewe2}
F. Riewe: Phys. Rev. E \textbf{55} (1997) 3581.
\bibitem{klimek1}
M. Klimek: Czech. J. Phys. \textbf{51} (2001) 1348.
\bibitem{klimek2}
M. Klimek: Czech. J. Phys. \textbf{52} (2002) 1247.
\bibitem{klimek3}
M. Klimek: Czech. J. Phys. \textbf{55} (2005) 1447.
\bibitem{agrawal}
O. P. Agrawal: J. Math. Anal. Appl. \textbf{272} (2002) 368.
\bibitem{Dreisigmeyer03}
D. W. Dreisigmeyer and P. M. Young: J. of Phys. A: Math. Gen.
\textbf{36} (2003) 8297.
\bibitem{Dreisigmeyer04}
D. W. Dreisigmeyer and P. M. Young: J. of Phys. A: Math. Gen.
\textbf{37} (2004) L117.
\bibitem{dumitru2}
D. Baleanu: \emph{Constrained systems and Riemann-Liouville
fractional derivative}, Proceedings of \emph{$1^{st}$ IFAC
Workshop on Fractional Differentiation and its Applications}(
Bordeaux, France, July 19-21, 597 (2004)).
\bibitem{Agrawalcankaya1} O. P. Agrawal: Generalized Euler-Lagrange
equations and transversality conditions for FVPs in terms of
Caputo Derivative, \emph{in Proc. MME06, Ankara, Turkey, April
27-29, 2006} (Eds. K. Tas, J.A. Tenreiro Machado and D. Baleanu),
to appear in J. Vib. Contr. (2006).
  \bibitem{eqab}
E. M. Rabei, K.I.  Nawafleh, R.S. Hijjawi, S. I. Muslih and D.
Baleanu: J. Math. Anal. Appl., in press (2006).
\bibitem{ubooks1}
D. Baleanu and S. Muslih: New trends in fractional quantization
method.{\it Intelligent Systems at the Service of Mankind}, volume
2. U-Books Verlag, Augsburg, Germany, 1st edition, December
(2005).

\bibitem{ubooks2}
D. Baleanu: Constrained systems and Riemann-Liouville fractional
derivatives.{\it Fractional differentiation and its applications}.
U-Books Verlag, Augsburg, Germany, November (2005).
\bibitem{dumitrusami}
S. I. Muslih and D. Baleanu: J.  Math. Anal.Apl. \textbf{304}
(2005) 599.
\bibitem{sp}
D. Baleanu:  Signal Processing \textbf{86(10)} (2006) 2632.

\bibitem{dsa}
S. Muslih and D. Baleanu: Czech. J. Phys. \textbf{55(6)} (2005)
633.
\bibitem{dsb}
D. Baleanu and S. Muslih: Physica Scripta \textbf{72(2-3)} (2005)
119.

\bibitem{Agrawalcankaya2} O. P. Agrawal: A Formulation and a Numerical Scheme for Fractional Optimal Control Problems, \emph{in
Proceedings of the IFAC/FDA06, Porto, Portugal, July 19-21,
(2006)}.
\bibitem{agrawal3}
O. P. Agrawal: Nonlinear Dynamics \textbf{38} (2004) 323.
\bibitem{Agrawalcankaya3}
O. P. Agrawal and D. Baleanu: A Hamiltonian Formulation and a
Direct Numerical Scheme for Fractional Optimal Control Problems,
\emph{in Proc. MME06, Ankara, Turkey, April 27-29, 2006} (Eds. K.
Tas, J.A. Tenreiro Machado and D. Baleanu), to appear in J. Vib.
Contr. (2006).

\end{thebibliography}
\end{document}